\documentclass[12pt,preprint]{aastex}
                
\usepackage{color}
\usepackage{times}
\usepackage{graphicx}
\usepackage{amssymb}
\usepackage{epstopdf}

\usepackage{times}

\usepackage{color}
\definecolor{aargh}{rgb}{0.5,0.2,0.8}

\usepackage{color}
\definecolor{red}{rgb}{1.0,0.0,0.0}
\newcommand{\ACBc}[1]{{ \color{black}{#1}}}

\newcounter{lastnote}

\newcommand{\microjy}{\mu \rm Jy} 
\newcommand{\mjy}{\rm mJy} 
\newcommand{\pb}{\rm beam^{-1}} 

\begin{document}

\title{Constraining the Planetary System of Fomalhaut Using High-Resolution ALMA Observations}
\author{A.~C.~Boley\altaffilmark{1}\altaffilmark{2}, M.~J.~Payne\altaffilmark{1},
S.~Corder\altaffilmark{3}, W.~Dent\altaffilmark{4},
E.~B.~Ford\altaffilmark{1}, and M.~Shabram\altaffilmark{1}}

\altaffiltext{1}{Department of Astronomy, University of Florida, 211 Bryant Space Science Center, Gainesville, FL, 32611}
\altaffiltext{2}{Sagan Fellow}
\altaffiltext{3}{North American ALMA Science Center, National Radio Astronomy Observatory, 520 Edgemont Road, Charlottesville, VA, 22903}
\altaffiltext{4}{ALMA, Alonso de Cordova 3107, Vitacura, Santiago, Chile}

\begin{abstract}\ACBc{The dynamical evolution of planetary systems leaves observable signatures in debris disks.}
Optical images trace micron-sized grains, which are strongly affected by stellar radiation and need not coincide with their parent body population.  
\ACBc{Observations of mm-size grains accurately trace parent bodies, but previous images lack the resolution and sensitivity needed to characterize the ring's morphology.}  
Here we present ALMA 350 GHz observations of the Fomalhaut debris ring. 
These observations demonstrate that the parent body population is 13-19 AU wide with a sharp inner and outer boundary. 
We discuss three possible origins for the ring, and suggest that debris confined by shepherd planets is the most consistent with the ring's morphology. 
\end{abstract}


\maketitle

\section{Introduction}
 The Fomalhaut debris system is a natural laboratory for testing planet formation theories.  
The nearby \citep[7.69 pc,][]{perryman_1997_AnA_323} A3V star \citep{difolco_2004_AnA_426} is surrounded by an eccentric debris ring with a peak brightness in scattered optical light at a semi-major axis $a\sim 140$ AU \citep[][henceforth K05]{kalas_etal_2005_nature_435}. 
The inner edge is sharply truncated, which, along with the ring's eccentricity, suggests that a planet is shaping the ring's morphology  \citep{wyatt_etal_1999_apj_527,quillen_2006_mnras_372,chiang_etal_2009_apj_693}.   
A candidate for the reputed planet (Fom b) has been discovered in the optical \citep{kalas_etal_2008_science_322}. 
The observations are consistent with a super-Earth mass planet embedded in a planetesimal swarm \citep{kennedy_wyatt_2011_mnras_412}.
 
Independent of direct detection of any planet,  observations of the ring's morphology can constrain the dynamical history of the Fomalhaut system, including properties of any planets near the ring.  
The effect of radiation pressure on a dust grain's orbital eccentricity, assuming an initially circular orbit, is $e=(2 \rho_s s/(\rho_{\infty} s_{\infty})-1)^{-1}$, where $\rho_s$ is the grain's internal density and $s$ the grain radius. 
\ACBc{Here, ${\infty}$ represents grains that are unbound by radiation pressure. 
For Fomalhaut, $s_\infty\sim 8\micron$ for $\rho_\infty\sim 1 \rm~g/cc$.
Most of the scattered optical light should be from the smallest bound grains \citep[between $\sim 8$ and $16\micron$, see ][]{chiang_etal_2009_apj_693}. 
The expected free eccentricity of a  $16\micron$ grain due to radiation pressure is $\sim 0.3$ (for $\rho_s=\rho_\infty$), while grain sizes near 1 mm will have free eccentricities $e<0.01$, making mm grains excellent tracers of parent bodies.
While previous observations of mm grain emission do detect large grains, they lack the resolution needed to characterize the parent body morphology} \citep{holland_etal_1998_nature_392,ricci_etal_2012_arxiv}.
Here, we present high-resolution 850 $\micron$ ALMA images that resolve the northern section of Fomalhaut's ring. 

\section{Observations and Reduction:}

Fomalhaut's ring was observed using ALMA cycle 0 in the compact configuration, measuring projected baselines from 14 to 175m (Table \ref{table:obs}). 
\ACBc{Observations were} centered on the expected position of Fom b at RA = 22h:57m:38.65s and $\delta$ = -29d:37':12.6'' (J2000, proper motion included).   
The total on-source integration time was 140 min. 
The observations were taken at 
357 and 345 GHz (in the upper and lower sidebands) using the Frequency Domain Mode in dual polarization with $4\times $1875 MHz bandpasses (2 in each of the sidebands). 
Neptune was used as an absolute flux calibrator, and J1924-292 for bandpass calibration. 
Atmospheric variations at each antenna were monitored continuously using Water Vapor Radiometers (WVR), as well as regular hot/ambient load measurements. 
For time-dependent gain calibration, the nearby quasar J2258-279 was observed every 8 minutes. 
Data were reduced using CASA 3.4 \citep{mcmullin_2007_aspc_376}: calibration involved removing the effects of rapid atmospheric variations at each antenna using the WVR data, correcting the time- and frequency-dependence of system temperature, and correcting for the complex antenna-based bandpass and time-dependent gain. 
Amplitude calibration used the CASA Butler-JPL-Horizons 2010 model for Neptune, which gives an estimated systematic flux uncertainty of $\pm$10\%. 
The calibrated measurement set was spectrally binned to a channel spacing of 49MHz, and then CLEANed using the Cotton-Schwab algorithm, combining all channels to give the final continuum image.
The primary beam correction \ACBc{(PBC)} was performed using the voltage pattern for an Airy function in with an effective dish diameter of 10.7 m, a blockage diameter of 0.75 m, a maximum radius of $1.784^{\circ}$, and a reference frequency of 1 GHz \ACBc{(see CASA functions vp.setpbairy and sm.setvp).}

\begin{table}
\begin{center}
\caption{Observing Log: Target Fomalhaut's Ring and Fomalhaut b. The precipitable water vapor  (PWV) is given in mm. \label{table:obs}}
\begin{tabular}{llccc}\hline
Date & Time  & On-Source  & Antennas & PWV\\
 dd/mm/yyyy    & (UTC)&(min)&     & \\\hline
22/09/2011 & 23:45:00.0 &  25 & 13 & 0.63 \\
23/09/2011 & 00:49:30.0 & 25 & 13 & 0.48\\
23/09/2011 & 02:08:15.3 & 25 & 13 & 0.80\\
18/10/2011 & 03:31:00.9 & 25 & 15 & 0.62\\
18/10/2011 & 04:42:55.6 & 20 & 15 & 0.63\\
18/10/2011 & 05:44:40.4 & 20 & 15 & 0.57\\
\end{tabular}
\end{center}
\end{table}

\section{Results}

\begin{figure*}[t]
\begin{center}
\includegraphics[width=14cm]{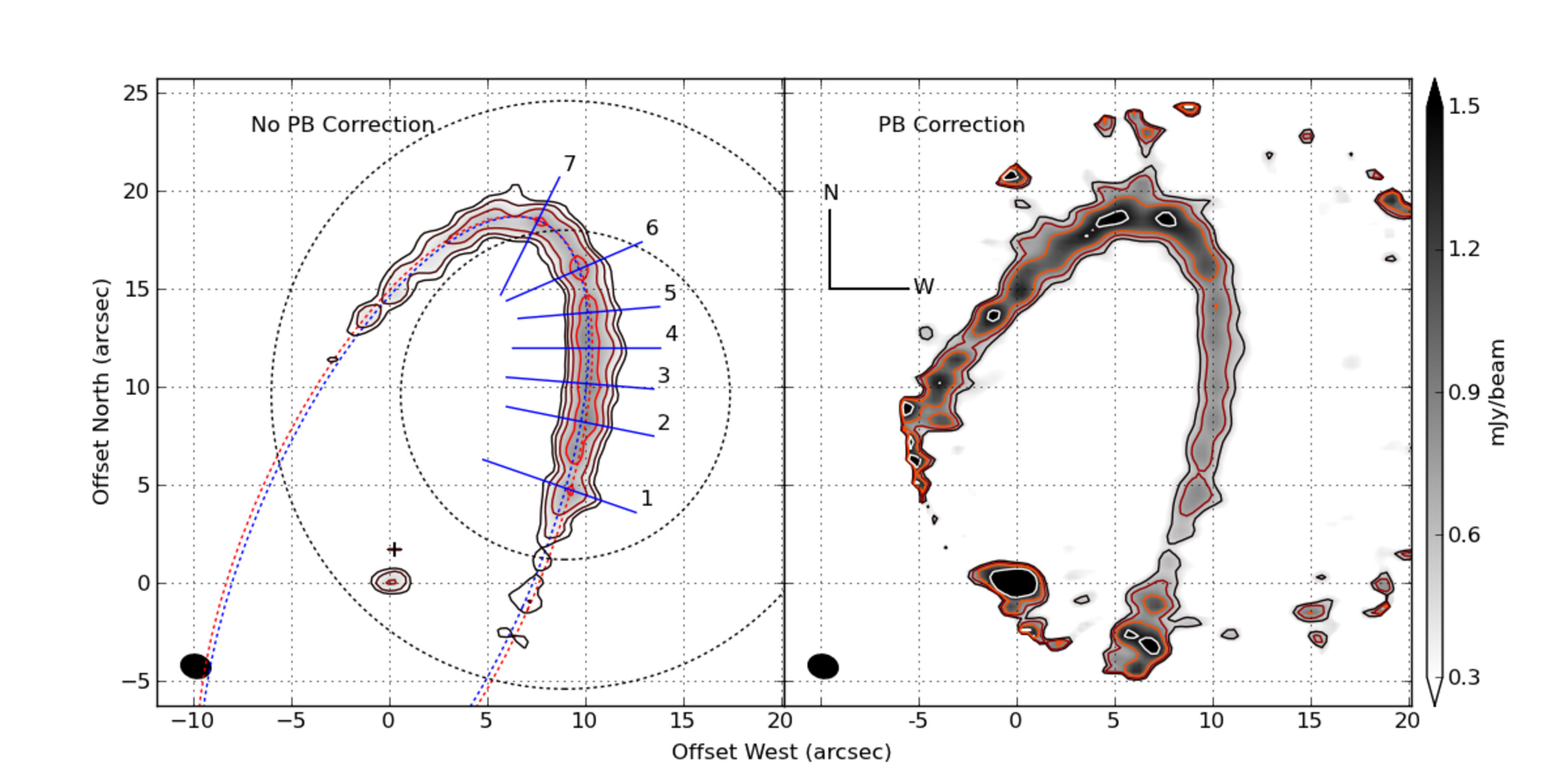}\\
\includegraphics[width=14cm]{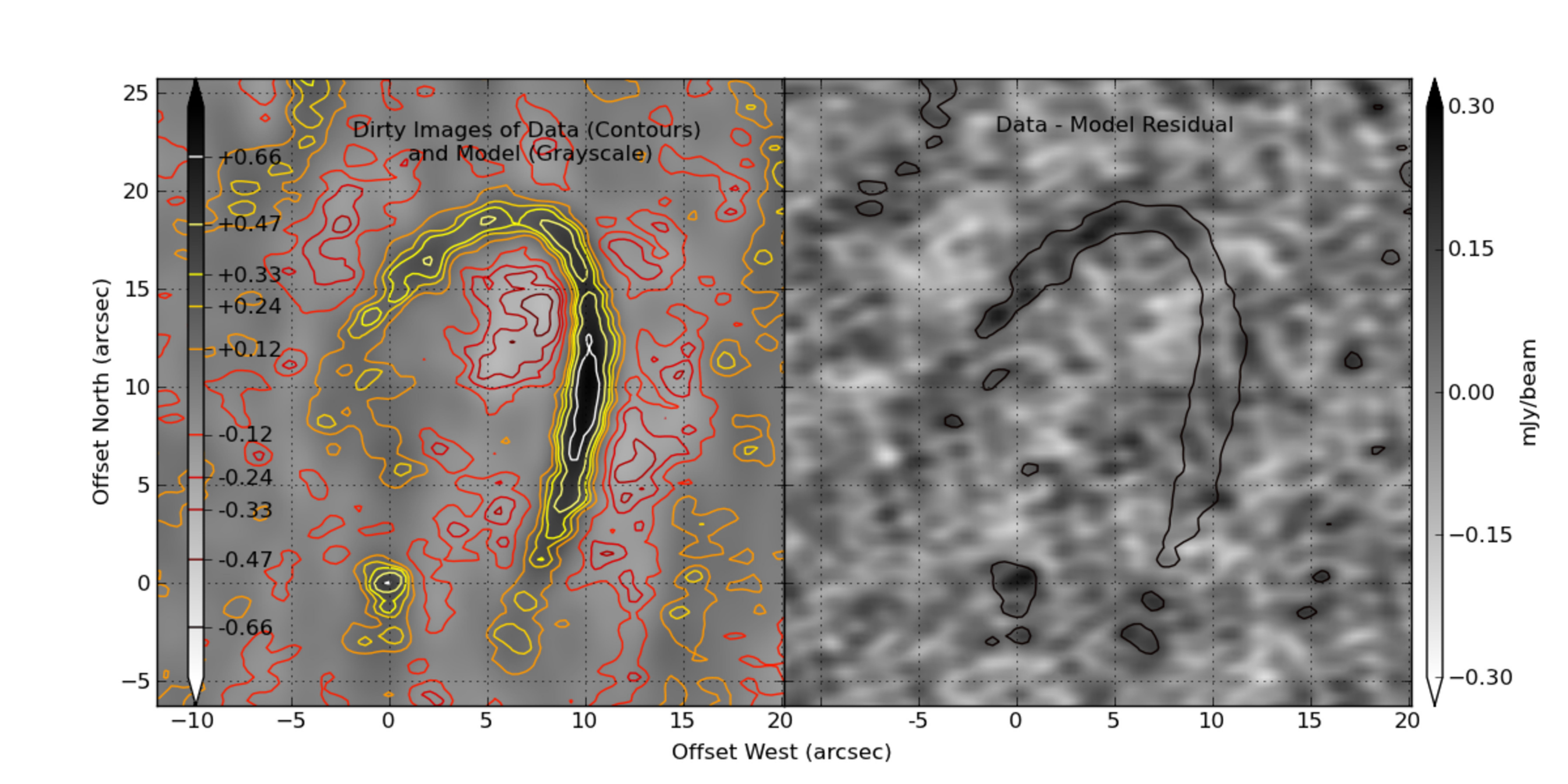}
\end{center}
\caption{
{\it Top-left:} 350 GHz ALMA image of Fomalhaut's ring.  
The RMS noise is $\sim 60\ \microjy~\pb$, and the contours represent 0.24, 0.33, 0.47, 0.66, 0.92, and 1.4 $\mjy~\pb$.
The circles are centered on the pointing center, and their diameters show the beam's half and 7\% power. 
Ellipses with ${\rm PA}=336^{\circ}$, and $i=66^\circ$ (red) and $67^{\circ}$ (blue) are shown, with the ellipse centers given by the plus sign. 
The slices labeled 1-7 are used to show surface brightness profiles in Figure 2.
The coordinate axis is centered on the star, which has a peak brightness of $0.49\ \mjy~\pb$ (uncorrected).
{\it Top-right:} The primary beam corrected image.
The ring becomes bright near the {\it ansa}, and remains bright through the inferred apocenter. 
The contours are the same as in the top-left, but begin at 0.47 $\mjy~\pb$.  
{\it Bottom-left:} The best-fit skymodel is compared with the dirty ALMA image. 
The grayscale and contours show $\mjy~\pb$, with the contour levels labeled on the colorbar. 
{\it Bottom-right:} The residual of the data minus the model.  
The model subtracts the emission well, but there remains excess emission in the NE and around the star. 
\label{fig:image}
}
\end{figure*}

In Figure 1 we present the cleaned and primary beam-corrected images. We refer to these as the uncorrected and corrected images, respectively.      
In the uncorrected image, the peak brightness is $\sim 0.84\ \mjy~\pb$, the total flux density of the image is 20.5 mJy, and the RMS is $\sim60\ \microjy~\pb$ (0.32 mK)\footnote{Unless otherwise stated, relative errors are assumed to be equal to the RMS value of $60 \mu\rm Jy~beam^{-1}$. Flux densities are only expected to be good to about 10\%.}.  \ACBc{The synthesized beam is $\sim1.5''\times 1.2''$ with position angle $77^{\circ}$.}
 
The corrected image shows all emission to 7\% power, which includes the ring and all apparent emission related to the star.  
The total flux in the corrected image is 45.5 mJy.
The projected image accounts for approximately half of the ring, so assuming equal brightness for the second half and accounting for the expected contribution from the star (3 mJy), we find a total flux density of $\sim 85$ mJy, reasonably consistent with previous measurements \citep[$81\pm7.2$ and $97\pm 5$ mJy][]{holland_etal_1998_nature_392,holland_etal_2003_apj_582}.  
{\it Almost all of the emission is confined to a thin ring.}

The ring's surface brightness reaches a maximum of $1.54\ \mjy~\pb$ near the {\it ansa}\footnote{{\it Ansa} refers to the sky-projected section of an astrophysical ring that is farthest from its central body, tradiationally referring to planetary rings.} and remains bright through the inferred apocenter.  
There also appears to be a reduction in the ring brightness to the SW.
We will discuss the significance of these azimuthal variations below.
The star's peak brightness is $3.4\ \mjy~\pb$ and has a flux density $\sim 4.4$ mJy (3 mJy expected \ACBc{for a blackbody photosphere}), which suggests possible excess emission.  
\ACBc{We caution that the measurement of total stellar flux is strongly dependent on the PBC. Nonetheless, a detection of excess emission is consistent with the results of \citep{stapelfeldt_etal_2004_apjs_154}. 
The PBC error would need to be 47\% at the 10\% power point to account for the excess, more than 4 times the ALMA beam specification which has yet to be completely verified.} 

Without the second {\it ansa}, we cannot reliably determine the ellipse. 
We show in Figure 1 two ellipse segments that are based on fits to optical images (K05).
The center for both ellipses, denoted by the plus sign, is 0.29'' W and 1.7'' N of the star. 
The ellipses have a \ACBc{semi-major axis} $a=18.31$'' and a position angle (${\rm PA}=336^{\circ}$). 
Assuming the ellipses represent sky-projected circles, the red ellipse has an inclination of $i=66^\circ$ (K05) and the blue $67^\circ$, which more closely corresponds to our best-fit sky models (below). 

Surface brightness profiles of the ring (Fig.~2) are taken along the 7 slices shown in Figure 1 using bilinear interpolation.
Each slice can be fit by a single Gaussian except slice 4, which requires a double Gaussian due to elongated emission.  Altogether the slices, excluding slice 4, have a combined FWHM $\sim~2.19"$, corresponding to a projected parent body width $\sim17$ AU, well-resolved by the synthesized beam. 

Deprojected slices through the corrected and uncorrected images at the ring's {\it ansa} are shown in Figure~2, which  includes all points within $\pm 2^\circ$ of the PA.
The ${\rm FWHM}= 1.87\pm0.03''$ and $1.85 \pm 0.03''$ for the corrected and uncorrected profiles, respectively.  
The surface brightness in the corrected image peaks at $\sim 18.4''$ (141.5 AU).
The peak is similar to that found for the small grains seen in the optical light radial brightness profile (K05), but the mm grains are much more tightly confined, consistent with radiation pressure effects. 
 
We estimate the mm grain mass by assuming the ring is optically thin and comprised of grains with $s\sim 1$ mm and $\rho_s\sim2.5$ g/cc.
The grain temperature $T_g\approx 0.7 T_{\rm star} (R_{\rm star}/D)^{1/2}$, where $T_{\rm star}=8750$ K and $R_{\rm star}=1.82 R_{\odot}$ are the stellar temperature and radius \citep{difolco_2004_AnA_426}, respectively.   
At $D=140$ AU, $T_g\approx 48$ K.
The ring's total flux density of 82 mJy (excluding the expected 3 mJy from the star), so $M_{\rm mm}\sim 0.017~{\rm M}_{\oplus}$. 

If collisions are the main removal mechanism of mm grains, then a lower limit can be placed on the parent body mass
$M_{\rm pbdy}/t_{\rm age} \sim M_{\rm mm}/t_{\rm coll}$  for ring age $t_{\rm age}$ and mm-grain collision timescale $t_{\rm coll}\sim (\tau \Omega)^{-1}$. 
The vertical optical depth through the ring can be estimated by $\tau\sim \rm (surface~brightness/flux~density~per~grain)(\pi s^2)$.
The peak of the ring's surface brightness is $\sim 1\ \mjy~\pb$. 
The synthesized beam effective area is $\sim 2~{\rm AU}^2$ at 7.69 pc.  
For $s\sim 1$ mm and $T_g\sim 48 K$, $\tau\sim 10^{-4}$. 
At 140 AU, $t_{\rm coll}\sim 2$ Myr. 
Combined with the above mass estimate, the ring should be losing mass at a rate of $\dot{M}\sim 0.01 M_{\rm Earth}$ $\rm Myr^{-1}$, \ACBc{requiring $M_{\rm pbdy}>1.7 M_{\rm Earth}$ (for $t_{\rm age}=200$ Myr), consistent with \citep{chiang_etal_2009_apj_693}.}

\ACBc{
Best-fit models for the spatial distribution of mm grains are selected by minimizing the difference between model and ALMA visibilities, where the $u$-$v$ plane sampling from the observations is used in CASA to observe each model (tool sm.predict).  
Ring models are produced using the following assumptions: (1) The ring is circular and has an offset relative to the star that is in agreement with the deprojected ellipse models (see Fig.~1), approximating a grain orbital eccentricity $\sim 0.1$. 
(2) The ring is uniform in azimuth.  
(3) The radial profile for mm grains is described by either a Gaussian distribution in semi-major axes or a power law in surface density.  For the Gaussian, the FWHM is varied from 9.4 to 21 AU, and the inner radial half-maximum point from 131 to 140 AU.
For power law surface density profiles, the annular width is varied from 10 to 30 AU, the inner edge from 133 to 138 AU, and the power law index from -9.5 to -1.5.   
(4) The vertical distribution for grains follows an exponential decay with a scale height given by a constant angle above the midplane between 0 and 2.5$^{\circ}$. 
In each model, the emission from grains at a given distance from the star is calculated from the Planck function with the grain temperature $T_g$ determined as described above. 
The total flux density for the ring at 850$\micron$ is normalized to either 80, 90, or 100 mJy.
The star's flux is set to 3 mJy.
Each model is projected onto the sky with ${\rm PA}=336^\circ$ and $i=66$ to $67.5^{\circ}$.  
The model ring's apocenter is set to the inferred apocenter of Fomalhaut's ring.  
It should be noted that models of a uniform ring can become brighter near the {\it ansae} due to projection effects.  
Smooth, symmetric models cannot, however, account for the excess brightness near apocenter. 
Such brightening could be explained by variations in the mass distribution of grains or by variations in ring's radial thickness. Collisions between parent bodies could also produce asymmetries \citep{wyatt_dent_2002_MNRAS_334}.
}

The best-fit models with a Gaussian \ACBc{semi-major axis} distribution have a scale height corresponding to an opening angle of $1.0^{\circ}\pm 0.25$, a ${\rm FWHM} =16 \pm 3$ AU, and an inner radial half-maximum at $135^{+0.5}_{-1}$ AU. \ACBc{The $1$-$\sigma$ errors for model parameters are estimated by including models for which $\Delta \chi^2 < 1$.}
Models with $i=66.75^{\circ}$ and a total flux density of 80 mJy are preferred.
Most of the dimming in the SW is consistent with a loss of sensitivity (Fig.~1). 
The appearance of a hole may be due to brightening in the southern-most part of the ring, which may be due to noise.
The NE is not subtracted in the residual image (Fig.~1, bottom-right) and contains contiguous excess surface brightness.  
The star is also not well-subtracted, suggesting possible excess.  
Some of the extended surface brightness noted in slice 4 of Fig.~1 remains, but is consistent with a high-noise peak. 

Some select, steep power law surface density profiles are also consistent with the data.  
These are limited to a surface density $\propto r^{-8.5}$,  an inner edge of $135^{+1}_{-1.5}$ AU, an opening angle $1.0^{\circ}\pm 0.5$, and a preferred $i=67.25^{\circ}$.  
For these power laws,  the half-maximum width is about 11.4 AU.  

\begin{figure}[t]
\includegraphics[width=8cm]{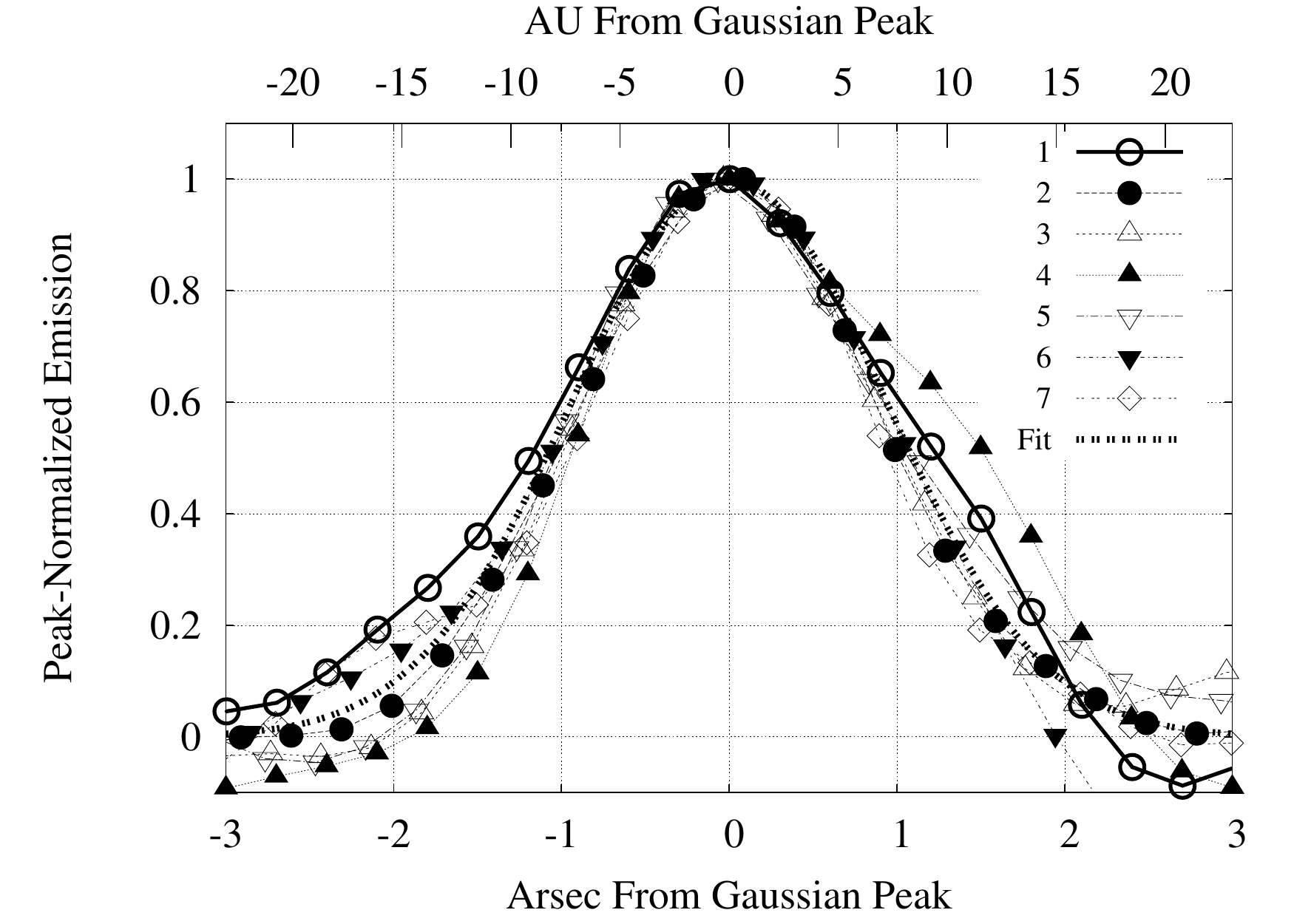}
\includegraphics[width=8cm]{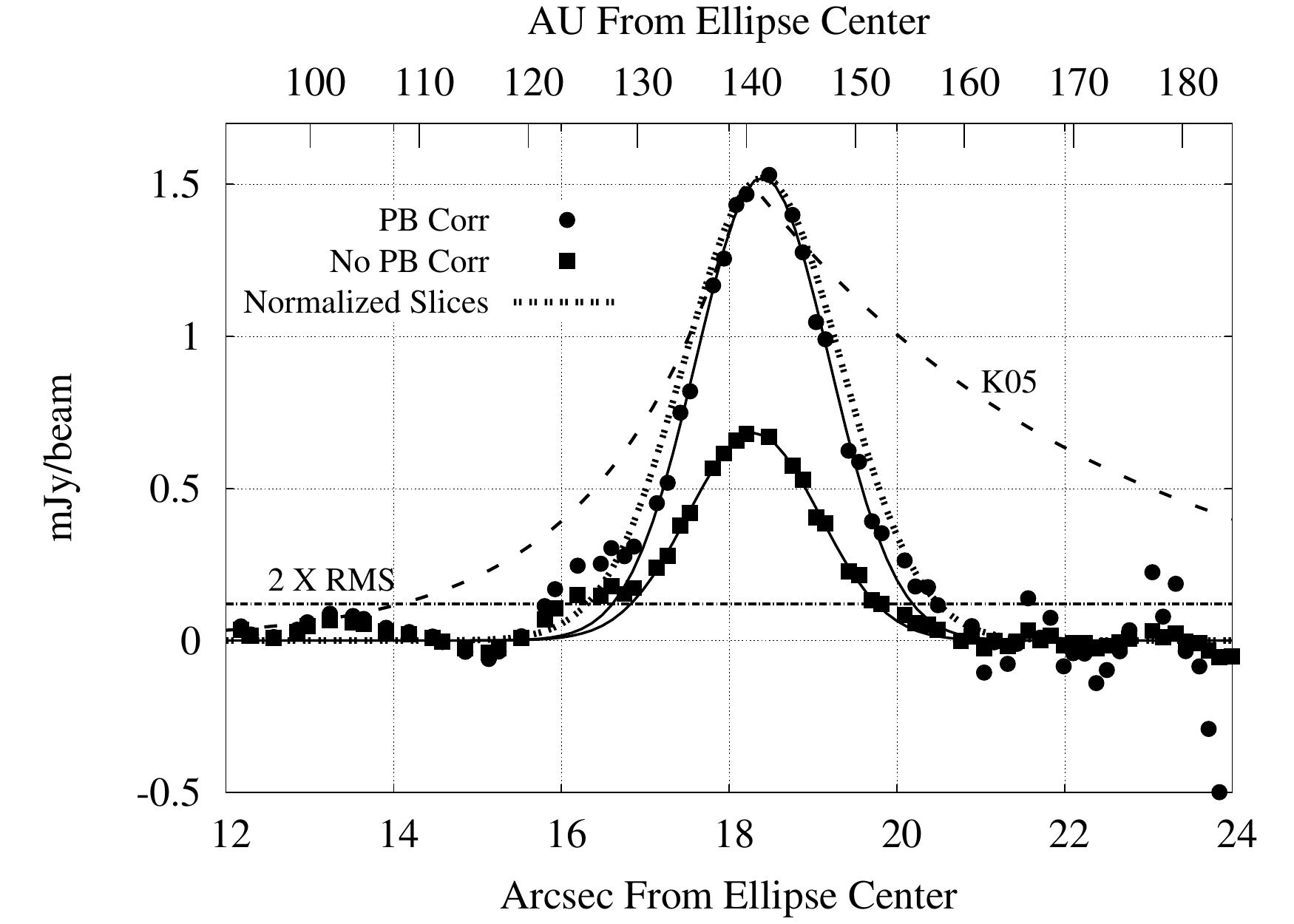}
\caption{
{\it Left:}  Normalized surface brightness profiles for slices 1-7 in Fig. 1. All profiles except slice 4 can be fit by a single Gaussian. A fit through all slices, excluding slice 4, is shown with the \ACBc{dot-dot} curve. 
 {\it Right:}  The deprojected radial surface brightness profile through the {\it ansa} for the corrected and uncorrected data. 
Black curves show corresponding Gaussian fits. 
The heavy-dashed curve shows the scaled scattered optical light profile (K05).
The \ACBc{dot-dot} curve shows the scaled Gaussian fit for the combined slices in the left panel. \label{fig:deproject}}
\end{figure}

\section{Discussion}

Any formation mechanism for Fomalhaut's ring must address the following constraints: 
(a) The ring has an eccentricity $\sim 0.1$.  
(b) The vertical scale height of the parent bodies is consistent with an opening angle of $\sim 1.0^{\circ}$ from the midplane. 
(c) The FWHM of the parent bodies for the Gaussian model is $\sim 16$ AU, giving a width-to-height aspect ratio $\sim 7$. 
(d) The outer edge of the parent bodies is consistent with being as sharply truncated as the inner edge. 
(e) The micron grains are more radially extended than the parent bodies, but this is satisfied for a variety of models due to radiation pressure.  
We consider three formation models.
One is that the ring is sculpted by a single interior planet \citep{wyatt_etal_1999_apj_527,kalas_etal_2005_nature_435,quillen_2006_mnras_372,kalas_etal_2008_science_322,chiang_etal_2009_apj_693}.
To account for requirement (d), this solution requires an abrupt truncation of the outer disk through, for example, a stellar flyby \citep{ida_etal_2000_apj_528}. 
We will return to this scenario below.

Another possible model is that the ring is a remnant from a collision between two planets \citep{mamajek_meyer_2007_apj_668}.
Unless constrained, the radial width of the ring will slowly spread due to collisions between ring particles with different semi-major axes. 
Take the vertical velocity dispersion $\delta v_z \sim v_K \sin i $ for Keplerian orbital speed $v_K$. 
If the radial velocity dispersion is of the same order, then the typical $a$ difference between two colliding particles is $\delta a\sim a \delta e$ to factors of order unity, which for free eccentricity $\delta e\sim \sin i$ is the scale height of the disk, $h$.  
The effective collisional viscosity of the ring $\nu_c \sim h^2/t_{\rm coll}$.
Using a scale height of 2.5 AU and a collisional time between mm grains of 2 Myr, $\nu_c \sim 2.2\times 10^{13}$ cm$^2$/s, which means the ring is diffusing on timescales $\sim 90$ Myr.  
If the ring is younger than the star, no planetary influence is necessary apart from the dynamics that led to the collision.
However, shattering gravitationally bound objects requires impact speeds several times the mutual escape speed ($v_{\rm esc}^2= 2 G (M_1+M_2)/(R_1+R_2)$, for bodies with masses $M_1$ and $M_2$ and radii $R_1$ and $R_2$). \citep{asphaug_2010_ChEG_70,leinhardt_2012_apj_745}.  
\ACBc{For two $1 M_{\rm Earth}$ planets ($M_{\rm pbdy} \gtrsim 1 M_{\rm Earth}$ for $t_{\rm age}\sim 90$ Myr),  $v_{\rm esc}\sim 11$ km/s.
The orbital escape speed at 140 AU from Fomalhaut is only about 5 km/s.  
The impact speeds necessary to destroy Earths are not attainable.}

\begin{figure}
\centering
\begin{tabular}{c}
\includegraphics[angle=-90, width=16cm,trim = 0mm 0mm 0mm 0mm, clip]{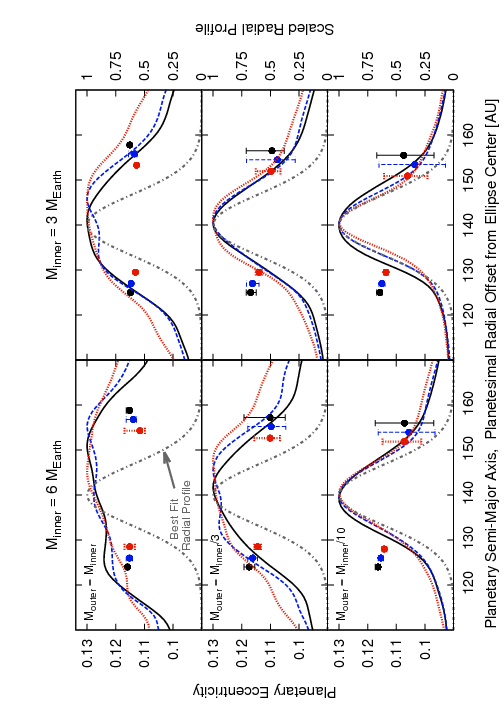}
\end{tabular}
\caption{
Radial profiles for a ring of parent bodies between two planets. 
Left-to-right: $M_{inner} = 6 M_{Earth}$ and $M_{inner} = 3 M_{Earth}$. 
Top-to-bottom: $M_{outer} = M_{inner}$, $M_{outer} = M_{inner}/3$, and $M_{outer} = M_{inner}/10$. 
\ACBc{Each color represents a profile-planet pair.
The gray dot-dashed line represents the best fit to the observed radial profile.}
The left y-axis shows the eccentricities of the planets, and the right shows the number surface density of particles, with each curve's peak normalized to unity.  
The range of eccentricities occupied by the planets is indicated by the vertical error bars.
The x-axis shows the semi-major axes for the planet and the radial distance of the planetesimals from the ellipse center.
\label{fig:NBODY2}}
\end{figure}

Finally, we propose a third formation model in which the ring's morphology is dominated by shepherd planets, analogous to the shepherd moons Cordelia and Ophelia of Uranus's $\epsilon$ ring \citep{goldreich_tremaine_1979_nature_277,smith_etal_1986_science_233}. 
Here, the ring is confined by angular momentum exchange between ring particles and the planets.  
We explore this possibility with N-body simulations, using the Bulirsch-Stoer algorithm in Mercury \citep{chambers_1999_mnras_304}.
In each simulation, two planets of mass $M_{inner}$ and $M_{outer}$ are placed at $a_{inner}$ and $a_{outer}$ \ACBc{relative to the $2.3\,M_{\odot}$ star}.  
Their relative inclinations are set to zero. 
Planet eccentricities are set to force the planetesimals  \citep[see method in][]{chiang_etal_2009_apj_693} at $\sim 143\,\rm AU$ to have the eccentricity observed for the ring ($\sim0.11$). 
For each run, $10^5$ massless particles are placed between $r_{inner}$ and $r_{outer}$ with a surface density profile $\propto r^{-1}$, where $r_{inner} = \texttt{max}(125 {~\rm AU},a_{inner}+1{~\rm AU})$ and $r_{outer} = \texttt{min}(160 {~\rm AU},a_{outer}-1 {~\rm AU})$. 
The integrations run for $10^8$ years. 
Massless particles are removed from simulations if the particle collides with a planet or if its radius from the star exceeds $10^5\,\rm AU$.

The masses of the planets set the steepness and skew of the ring's radial profile (Fig.~3).
Mutual perturbations will cause the system to evolve, which narrows parameter space by demanding that interactions between the planets do not destroy the ring. 
In the massless ring-particle limit, either both planets must be $<3 M_{\rm Earth}$, or they must have an extreme mass ratio.
The planets do not need to be in resonance.
A super-Earth and a Mars-mass planet produce the most narrowly-peaked Gaussian for the radial parent body distribution among the simulations presented here.  
As the masses for shepherd candidates are comparable to the {\it minimum} estimated parent body mass in the ring, self-gravity of the ring may play a role in the system's evolution. 
If the scattered optical light observations have detected Fom b, then the candidate could be the innermost shepherd, although there is a discrepancy  of $\sim10$ AU between Fom b's proposed $a$ and the inner shepherd in our models.

Fomalhaut's debris has some similarities to the Kuiper belt. 
Both systems are thin rings, $\Delta r/a\sim 0.1$, which seem to owe their morphologies, at least in part, to the presence of planets. 
There are nonetheless substantial differences.
The outer edge of the Kuiper belt is near the 2:1 resonance with Neptune \citep{trujillo_etal_2001_apj_554}.
For the outer edge of Fomalhaut's ring to be at the 2:1 resonance with
an interior planet, the planet would need to have an $a\sim 95$ AU.
To truncate the ring at this location, the planet would need to be \ACBc{$\gtrsim 3 M_J$} \citep{chiang_etal_2009_apj_693} and could have been decected by {\it Spitzer} \citep{janson_etal_2012_arxiv}.
If the ring had formed due to outward migration of an interior planet, we would expect particles to be trapped in resonances, like the Plutinos in the 3:2 resonance with Neptune. 
We detect only a single ring. 
The Kuiper belt has \ACBc{a vertical thickness given by} a dynamically hot and cold population with Gaussian widths $\sigma =17^{\circ}\pm3$ and $\sigma=2.2{^\circ}~{}^{+0.2}_{-0.6}$, respectively \citep{brown_2001_aj_121}. 
While the Kuiper belt object inclination distribution seems to be consistent with truncation due to a stellar flyby \citep{ida_etal_2000_apj_528}, Fomalhaut's parent body population may be too cold to be explained by this mechanism. 
This leads us to favor shepherd planets as the explanation for the ring's morphology.

\acknowledgements
We thank Paul Kalas \ACBc{and the referee for helpful comments} and ALMA NAASC for technical support.
A.C.B's support was provided under contract with the California Institute of Technology funded by NASA through the Sagan Fellowship Program. 
M.J.P.'s and E.C.B.'s contributions are supported by the National Science Foundation (NSF) under grant no.~0707203.
M.S.'s support was provided in part by the National Radio Astronomy Observatory (NRAO) through award SOSPA0-007.
The NRAO is a facility of the NSF operated under cooperative agreement by Associated Universities, Inc.
This paper makes use of the following data from the Atacama Large Millimeter/submillimeter Array
(ALMA): ADS/JAO.ALMA\#2011.0.00191.S. 
The Joint ALMA Observatory is a partnership of the European
Organization for Astronomical Research in the Southern Hemisphere, the National Astronomical
Observatory of Japan (on behalf of the National Institutes of Natural Sciences and Academia Sinica),
and the NRAO (managed by Associated Universities, Inc. on behalf
of the NSF and the National Research Council of Canada) in cooperation with the Republic of Chile.



\end{document}